\documentclass[12pt]{article}
\usepackage{graphicx}
\usepackage{latexsym}
\usepackage{amsmath,amsfonts,amsthm}
\newcommand{\AGf}{\ensuremath{\mathcal{A}}}
\newcommand{\PGf}{\ensuremath{\mathcal{P}}}
\newcommand{\RGf}{\ensuremath{\mathcal{R}}}

\begin{document}
\title{Statistics of lattice animals (polyominoes) and polygons}
\author{Iwan Jensen\thanks{e-mail: I.Jensen@ms.unimelb.edu.au}\mbox{ }
and Anthony J. Guttmann\thanks{e-mail: tonyg@ms.unimelb.edu.au}\\
Department of Mathematics and Statistics, \\
The University of Melbourne,\\
Victoria 3010, Australia}
\date{\today}
\maketitle
\bibliographystyle{plain}
\begin{abstract}
We have developed an improved algorithm that allows us to enumerate the 
number of site animals (polyominoes) on the square lattice up to size 46. 
Analysis of the resulting series yields an improved estimate,
$\tau =  4.062570(8)$, for the growth constant of lattice animals
and confirms to a very high degree of certainty that the generating
function has a logarithmic divergence. We prove the bound $\tau > 3.90318.$
We also calculate the radius of gyration of both lattice animals and 
polygons enumerated by area.
The analysis of the radius of gyration series yields the estimate
$\nu = 0.64115(5)$, for both animals and polygons enumerated by area.
The mean perimeter of polygons of area $n$ is also calculated.
A number of new amplitude estimates are given.
\end{abstract}

The enumeration of lattice animals is a classical combinatorial 
problem of great interest both intrinsically and as a paradigm
of recreational mathematics \cite{Golomb}. 
A lattice animal is a finite set of nearest neighbour sites on 
a lattice. The fundamental problem is the calculation of the number of
animals, $b_n$, containing $n$ sites. 
In the physics literature lattice animals are very often
called {\em clusters} due to their very close relationship
to percolation problems \cite{Stauffer}. Series expansions 
for various percolation properties, such as the percolation probability 
or the average cluster size, can be obtained from the perimeter polynomials. 
These in turn can be calculated by counting the number of lattice animals 
$b_{n,m}$ according to their size $n$ and perimeter $m$ \cite{Domb,Sykes}. 
Lattice animals have also been suggested as a model of branched 
polymers \cite{Lubensky}.
In  mathematics, and combinatorics in particular, the term 
polyominoes is frequently used. A polyomino is a set of lattice cells 
joined at their edges. So polyominoes are identical to site animals on the 
dual lattice. Furthermore, the enumeration of lattice animals has 
traditionally served as a benchmark for computer performance and algorithm 
design \cite{Martin}--\cite{Conway}. 

The enumeration of self-avoiding polygons is another classical
combinatorial problem {\cite{Hug95}. Most attention has been paid 
to the enumeration by perimeter, but enumeration by area is
an equally interesting problem. (For polyominoes, the ordinary
generating function of the number of polyominoes of perimeter $n$
has zero radius of convergence \cite{JGO00} and hence is of rather
less interest). Polygons enumerated by area are just the ``hole-free"
subset of polyominoes. There are exponentially fewer polygons
than polyominoes \cite{GJWE00}, but on universality grounds
one would expect the exponent $\nu$ characterising the radius of
gyration to be the same for polyominoes and polygons enumerated
by area. We confirm this expectation.

An algorithm for the calculation of $b_{n,m}$ has been published by
Martin \cite{Martin} and Redner \cite{Redner}. It was used by Sykes
and co-workers to calculate series expansions for percolation problems
on various lattices. In particular Sykes and Glen \cite{Sykes} calculated
$b_{n,m}$ up to $n=19$ on the square lattice, and thus obtained the number 
of lattice animals, $b_n=\sum_m b_{n,m}$, to the same order. Redelmeier 
\cite{Redelmeier} presented an improved algorithm for the enumeration of 
lattice animals and extended the results to $n=24$. This algorithm was 
later used by Mertens \cite{Mertens90} to devise an improved algorithm for 
the calculation of $b_{n,m}$ and a parallel version of the algorithm 
appeared a few years later \cite{Mertens92}. The next major advance
was obtained by Conway \cite{Conway} who used the finite lattice
method with an associated transfer-matrix algorithm to calculate
$b_{n}$ and numerous other series  up to $n=25$ \cite{CG95}. 
In unpublished work Oliveira e Silva
\cite{Silva} used the parallel version of the Redelmeier algorithm
\cite{Mertens92} to extend the enumeration to $n=28$. In this work
we use an improved version of Conway's algorithm to extend the enumeration 
to $n=46$. We also calculate the 
the radius of gyration of lattice animals and square lattice
polygons enumerated by area up to $n=42$. Further, we also calculate
the mean perimeter of polygons of area $n.$ Instructions for the
electronic retrieval of these series can be found at the end of this
Letter.

The method we use to enumerate site animals and polygons on the square lattice 
is based on the method used by Conway \cite{Conway} for the calculation of 
series expansions for percolation problems, and is similar to methods
devised by Enting for enumeration of self-avoiding polygons \cite{Enting}
or the algorithm used by Derrida and De Seze
to study percolation and lattice animals \cite{DeD82}. 
The number of animals that span rectangles of width $W$ and length $L$ 
are counted using a transfer matrix algorithm. A detailed description 
of the algorithm  will appear elsewhere \cite{Jensen00}.

The quantities and functions we consider in this Letter are:
(i) the number of polygons of area $n$, denoted $a_n,$ and the associated 
generating function, $A(y)= \sum a_n y^n$; (ii) the number of lattice 
animals $b_n$ and the associated generating function, 
$\AGf (u)= \sum b_n u^n$;  (iii) the first moment of the number 
$p_{n,m}$ of polygons of perimeter $m$ and area $n,$ 
$a_n\langle p \rangle_n = \sum_m mp_{n,m}.$ Then $\langle p \rangle_n $
is the mean perimeter of polygons of area $n.$
(iv) the mean-square radius of gyration of animals of area $n$, 
$\langle R_{\rm a}^2 \rangle _n$. 
(v) the mean-square radius of gyration of polygons of area $n$, 
$\langle R_{\rm p}^2 \rangle _n$. 
These quantities are expected to behave as

\begin{eqnarray}\label{eq:coefgrowth}
a_n & = & A \kappa^n n^{-1}[1+o(1)], \nonumber \\
b_n & = & B \tau^n n^{-1}[1+o(1)], \nonumber \\
a_n\langle p \rangle_n &=& AC\kappa^n[1+o(1)], \nonumber \\
\langle R_{\rm a}^2 \rangle_n & = & D n^{2\nu_{\rm a}}[1+o(1)], \nonumber \\
\langle R_{\rm p}^2 \rangle_n & = & E n^{2\nu_{\rm p}}[1+o(1)], 
\end{eqnarray}
\noindent
where $\kappa$ is the reciprocal $y_c^{-1}$ of the critical 
point of the polygon area generating function, 
and $\tau$ is the reciprocal $u_c^{-1}$ of the critical 
point of the animal generating function. 
From numerical evidence  \cite{GJWE00} it is well-established that 
both the polygon area generating function
and the animal generating function have a logarithmic singularity,
hence the factor $1/n$ in the first two equations above. Similarly, it 
is generally believed \cite{EG90} that $\langle p \rangle_n \sim n,$
so the $n$-dependence vanishes to leading order in the third
equation above.

The series studied in this Letter have coefficients which 
grow exponentially, with sub-dominant term given by a critical exponent.
The generic behaviour is $G(x) =\sum_n g_n x^n \sim (1-x/x_c)^{-\xi},$ 
and hence the coefficients of the generating function 
$g_n \sim \mu^n n^{\xi-1}$, where $\mu = 1/x_c$. To obtain the singularity 
structure of the generating functions we first used the numerical method of 
differential approximants \cite{Guttmann89}. Combining the relationship (given 
above) between the coefficients in a series and the critical behaviour of 
the corresponding generating function with the expected behaviour 
(\ref{eq:coefgrowth}) of the mean-square radius of gyration 
yields the following expectation for the animal generating functions:

\begin{eqnarray}\label{eq:genfunc}
\AGf (u)& = &\sum_n b_n u^n = A(u)\log(1-\tau u), \\
\RGf^2_{\rm a} (u)& = &\sum_n b_{n}\langle R_{\rm a}^2 \rangle _{n}n^2 u^n =
    \sum_{n} r_n u^n \sim R(u)(1-\tau u)^{-(2+2\nu_{\rm a})}. 
\end{eqnarray}
\noindent
Thus we expect these series to have a critical point,
$u_c=1/\tau$, and as stated previously the  animal generating
function is expected to have a logarithmic singularity. Similar expressions 
hold for the corresponding polygon area generating function, though with a 
different growth constant $\kappa$. The radius of gyration series are 
expected to diverge with exponents $2+2\nu_{\rm a}$, and $2+2\nu_{\rm p}$ respectively, 
though as we have argued above, we expect the exponents to be equal.

Estimates of the critical point and critical 
exponent were obtained by averaging values obtained from second order 
$[L/N;M;K]$ inhomogeneous differential approximants.   
In Table~\ref{tab:analysis} we have listed the estimates obtained from 
this analysis. The error quoted for these estimates reflects the spread 
(basically one standard deviation) among the approximants. Note that these 
error bounds should {\em not} be viewed as a measure of the true error as 
they cannot include possible systematic sources of error. From this
we see that the animal generating function has a singularity at 
$u_c=0.246150(1)$, and thus we obtain the estimate, $\tau =  4.06256(2)$,
for the growth constant. The exponent estimates are consistent with the 
expected logarithmic divergence.
The central estimates of $u_c$ obtained from the radius of gyration series
are a little larger than, but nonetheless consistent with those
from the animal
generating function. From this analysis we see that this series has a 
divergence at $u_c$ with an exponent $2+2\nu_{\rm a} = 3.2840(8)$, and thus
$\nu_{\rm a}=0.6420(4)$.

Once the conjectured exact value of the exponent has been confirmed we can 
obtain an improved estimate for the critical point. In 
figure~\ref{fig:anicpex} we have plotted estimates for the exponent 
 vs the corresponding estimates for the critical point $u_c$ as 
obtained from second order approximants to the animal generating function. 
From this figure we see that, as the estimates for the exponent
approach 0, the estimates of
$u_c$ approach 0.2461497. From the spread among the 
approximants we obtain out final estimate $u_c=0.2461496(5)$, and thus
the growth constant $\tau = 4.062570(8)$. An earlier analysis, based
on shorter series \cite{GJWE00} gave $\tau = 4.062591(9).$
A similar analysis for
the lattice tree generating function is given in \cite{Jensen00}, 
and there it is found that $\nu = 0.64115(5).$
The estimates of $u_c$ obtained from differential approximants to
the radius of gyration 
of animals is consistently larger than the above final estimate and 
this is probably the reason the estimate for $\nu_{\rm a}$  is slightly
larger than
that obtained from trees. Further evidence of this can be found by
looking at figure~\ref{fig:anirgs} where we have plotted the
estimates of the exponent $2+2\nu_{\rm a}$ vs $u_c$ as obtained from the
radius of gyration series for animals. Clearly as $u_c$ decrease
so does the exponent and as $u_c$ approaches 0.2461497 the exponent
gets closer to the estimate $2+2\nu_{\rm a} = 3.2823(1)$ obtained \cite{Jensen00}
for lattice trees and
the discrepancy is thus largely resolved. For this reason we claim that 
the most precise estimate for $\nu$ is the one obtained \cite{Jensen00}
 from the tree series. Further evidence for this claim is given
in \cite{Jensen00}.

Using this value of the exponent $\nu$, and the estimate of $u_c$ cited
above, we repeated the amplitude analysis of the animal series cited
in \cite{GJWE00}. Now however we have 46 terms. In \cite{GJWE00} we
found evidence that $b_n \sim \tau^n/n[d_0 + d_1/n + 
d_2/n^{\Delta_2} + d_3/n^{\Delta_3} + \cdots]$
with $\Delta_n = n.$ With the longer series we in fact find that 
$\Delta_2 = 1.5$ and
$\Delta_3 = 2,$ with subsequent values of $\Delta_k$ presumably increasing
by $1/2.$ We note that this is consistent with the known
correction-to-scaling exponent for polygons enumerated by perimeter,
$\Delta = 1.5$ \cite{Nie82}. By fitting to this form we estimate
$d_0 = 0.316915(10),$ $d_1 = -0.276(2),$ $d_2 = 0.335(10),$ 
and $d_4 = -0.25(5).$
The errors quoted tacitly assume that the critical point is correct.
A similar analysis for the radius of gyration series displays evidence
of a similar confluent term, and we find the data can be effectively fitted
by the following asymptotic form:
$n^2 b_n\langle R_{\rm a}^2 \rangle_n \sim \tau^n n^{2\nu+1}
[e_0 + e_1/n + e_2/n^{\Delta_2} + e_3/n^{\Delta_3} + \cdots].$ 
That is to say, the same confluent exponent is
observed, and the amplitudes may be estimated as:
$e_0 = 0.0599(2),$ $e_1 = -0.190(8),$ and $e_2 = 0.5(1).$ 
The quality of the fit was less satisfactory than the
corresponding fit to the total number of animals, and only three amplitudes
can be quoted with any confidence. Combining these,
we find
$\langle R_{\rm a}^2 \rangle_n \sim  n^{2\nu}[f_0 + f_1/n + f_2/n^{3/2}]$
where $f_0 = 0.1890(12),$ $f_1 = -0.435(26)$ and $f_2 = 1.4(4).$

For square lattice polygons enumerated by area, the series to 42 terms
is given in \cite{GJWE00}. The differential approximants are summarised
in Table 2, and on the basis of these, and a subsequent analysis that
assumes that the critical exponent is zero, we estimated the
connective constant to be $\kappa = 3.97094397(9).$ An amplitude
analysis, also given in \cite{GJWE00} gave
$$a_n \sim \kappa^n/n[0.408105 - 0.547/n + 0.63/n^2 + o(1/n^2)].$$
Here there is no evidence of a correction-to-scaling term $\Delta=1.5,$
though there is some evidence of a weaker non-analytic correction, 
perhaps consistent with $\Delta=2.5.$

The differential approximant analysis for the radius of gyration series
is also summarised in Table 2, and displays similar features to
that for animals, just discussed.
Using the quoted  value for $\kappa,$ a biased differential approximant
analysis of the generating function
$$\RGf^2_{\rm p} (y) = \sum_n a_{n}\langle R_{\rm p}^2 \rangle _{n}n^2 y^n =
    \sum_{n} r_n y^n \sim R(y)(1-\kappa y)^{-(2+2\nu_{\rm p})}$$ 
similar to that described above for lattice animals, gave mainly
defective approximants, though almost all exponent estimates were
clustered around $2 + 2\nu_{\rm p} = 3.283$ or $\nu_{\rm p} = 0.6415.$ This is
very close to the estimate obtained for both polyominoes and lattice trees,
cited above. Accordingly, we conjecture that the three exponents are
the same, and we take the seemingly most precise value, $\nu = 0.64115$
as found for lattice trees, as our preferred value. 

Using this value of the exponent, and the estimate of $\kappa$ cited
above, we repeated the amplitude analysis {\em mutatis mutandis} 
described above in our analysis of animals.
A similar fit to the radius of gyration series also showed 
evidence of a correction-to-scaling term $\Delta=1.5,$ and we found:
$n^2 a_n\langle R_{\rm p}^2 \rangle_n \sim \kappa^n n^{2\nu+1}
[0.08488 - 0.457/n + 0.77/n^{1.5} + 0.3/n^{2} + \cdots].$ 
Errors in the amplitude estimates are expected to be
confined to the last quoted digit.
Combining these results we find
$\langle R_{\rm a}^2 \rangle_n \sim  n^{2\nu}[g_0 + g_1/n + g_2/n^{3/2}]$
where $g_0 = 0.2080,$ $g_1 = -0.840$ and $g_2 = 1.9.$

A similar analysis of the first moment series was also made, and again
we found evidence of a non-analytic correction-to-scaling term 
$\Delta=1.5.$ More precisely, we found
$$a_n\langle p \rangle_n \sim \kappa^n[0.75715 - 0.064/n 
+ 0.07/n^{3/2} + O(1/n^2)],$$
so that
$$\langle p \rangle_n \sim 1.8552n + 2.33 + 0.17/\sqrt{n}.$$

Finally, we used the series to derive improved rigorous lower
bounds for the growth constants of lattice animals and trees.
Using concatenation arguments, Rands and Welsh \cite{Rands} showed
that if we define a sequence $p_n$ such that

\begin{equation}
b_{n+1}=p_{n+1}+ p_{n} b_{2}+\ldots p_3 b_{n-1}+p_2 b_n,
\end{equation}
\noindent
and construct the generating functions 
\begin{equation}
\AGf^* (u) = 1 + \sum_{n=1}^{\infty} b_{n+1}u^n
\end{equation}
\noindent
and
\begin{equation}
\PGf (u) = \sum_{n=1}^{\infty} p_{n+1}u^n
\end{equation}
\noindent
then
\begin{equation}
\AGf^* (u) = 1+ \AGf^* (u)\PGf (u)
\end{equation}
\noindent
and  $\AGf^* (u)$ is singular when $\PGf (u)=1$. The coefficients in
$\PGf (u)$ are obviously known correctly to the same order $N=2W_{\rm max}-1$
as $\AGf^* (u)$. If we look at the polynomial $P_N$ obtained by truncating
$\PGf (u)$ at order $N$ then the unique positive zero, $1/\tau_N$, of
$P_N-1=0$ is a lower bound for $\tau$. Using our extended series we
find that $\tau \geq 3.903184\ldots$.

In conclusion,
radically extended series for animals and polygons enumerated by area have
been presented. Improved estimates of critical points and critical exponents
have been made. The area generating function of both polyominoes
and polygons is found to have a logarithmic singularity, while
the radius of gyration exponent was estimated to be
$\nu = 0.64115.$ This value merits some discussion.
Two dimensional lattice models frequently have rational
critical exponents with typically one or two digit numerators
and denominators. In this case the closest ``small'' rational fraction is
$\frac{109}{170},$ a startlingly uncompelling one! However as the
animal problem is not conformally invariant, we have no theoretical 
reason to expect a rational exponent, and our result certainly
doesn't suggest one. Earlier, less precise estimates of $\nu$ have
been given in \cite{DeD82}, wherein the estimate $\nu = 0.6408(3)$
was made, and more recently in \cite{You+Rensburg} the Monte Carlo
estimate for lattice trees,
$\nu = 0.642(10)$ is given. Several earlier, less precise Monte
Carlo and series estimates are also referenced there, and the
correction-to-scaling
exponent is also studied, and the estimate $\Delta = 0.65 \pm 0.20$ 
given. However for the problems of both polyominoes and polygons,
we find no evidence of any such singularity with exponent less than 1.
This is reminiscent of the situation for the enumeration of polygons
by perimeter, where for many years various methods of analysis
yielded estimates in the range $0.5$ to $1.5.$ Only with very long
series \cite{JG99} did it become clear that there were no such
terms with exponent less than 1, and that the long-standing prediction
of Nienhuis \cite{Nie82} that $\Delta = 1.5$ 
was completely correct. We suggest that something
similar is the case here.

We have also obtained a more precise lower bound, $\tau > 3.90318,$
on the polyomino connective constant.

Finally we should comment briefly on the amplitude estimates we
have made. These may be summarised, following the definitions in
\ref{eq:coefgrowth}, as $A=0.408105(10),$ $B=0.316915(10),$ $C=1.8552(10),$
$D=0.1890(12),$ and $E=0.2080(2).$
For polygon enumeration by perimeter, there are a number
of universal ratios \cite{CG93} known. In the case considered
here, where we enumerate by area, there are no published predictions.
Many analogous relations would not exist, as in the perimeter case they
depend on theorems following from conformal invariance. While
certain products and quotients are suggestive, none are sufficiently
compelling as to lead us to believe that they are worth publishing. We
rather highlight this as an open problem, for which we provide
useful test data.

\section*{E-mail or WWW retrieval of series}

The series for the generating functions studied in this Letter 
can be obtained via e-mail by sending a request to 
I.Jensen@ms.unimelb.edu.au or via the world wide web on the URL
http://www.ms.unimelb.edu.au/\~{ }iwan/ by following the instructions.

\section*{Acknowledgements}

Financial support from the Australian Research Council is 
gratefully acknowledged, as is useful discussion with John Cardy.

\clearpage

\begin{table}
\caption{\label{tab:analysis} Estimates for the critical point $u_c$ and 
exponents $1-\theta_{\rm a}$ and $1+\theta_{\rm a}+2\nu_{\rm a}$ obtained from second order 
inhomogeneous differential approximants to the series for the generating 
functions of lattice animals and their radius of gyration. Also listed are 
the corresponding estimates for polygons. 
$L$ is the order of the inhomogeneous polynomial.}
\begin{center}
\begin{tabular}{rllll}
\hline \hline
\multicolumn{5}{c}{Square lattice site animals} \\ \hline
\multicolumn{1}{c}{L} &
\multicolumn{1}{c}{$u_c$} &
\multicolumn{1}{c}{$1-\theta_{\rm a}$} &
\multicolumn{1}{c}{$u_c$} &
\multicolumn{1}{c}{$1+\theta_{\rm a}+2\nu_{\rm a}$} \\ 
\hline 
 0  & 0.246149987(43)& -0.000523(46)                
    & 0.246150539(87)& 3.28413(11) \\         
 2 & 0.24614992(14)& -0.00043(14)       
    & 0.24615046(10)& 3.28402(28) \\         
 4 & 0.24615007(15)& -0.00055(16)        
   & 0.24615037(22)& 3.28394(30) \\         
 6 & 0.24614999(24)& -0.00046(25)        
   & 0.24615068(16)& 3.28426(22) \\         
 8 & 0.24615001(15)& -0.00052(13)         
   & 0.24615067(25)& 3.28432(44) \\         
10 & 0.24614997(22)& -0.00044(28)    
   & 0.24615055(31)& 3.28417(56) \\   
\hline \hline      
\multicolumn{5}{c}{Square lattice polygons enumerated by area } \\
\hline 
\multicolumn{1}{c}{L} &
\multicolumn{1}{c}{$y_c$} &
\multicolumn{1}{c}{$1-\theta_{\rm p}$} &
\multicolumn{1}{c}{$y_c$} &
\multicolumn{1}{c}{$1+\theta_{\rm p}+2\nu_{\rm p}$} \\ 
\hline 
 0  & 0.251829311(24)& -0.000022(23)
    & 0.25183133(81)& -3.2847(13) \\
 2  & 0.251829340(20)& -0.000051(19)
    & 0.25183043(73)& -3.28439(88) \\
 4  & 0.251829349(52)& -0.000059(53)
    & 0.25183052(62)& -3.2846(12) \\
 6  & 0.251829314(12)& -0.000025(12)
    & 0.2518302(11)& -3.2841(16) \\
 8  & 0.251829320(18)& -0.000031(19)
    & 0.25183035(50)& -3.28432(62)  \\
 10  & 0.251829319(12)& -0.000029(12) 
     & 0.2518298(19)& -3.2836(28) \\
\hline \hline        
\end{tabular}
\end{center}
\end{table}

\begin{figure}

\begin{center}

\includegraphics[scale=0.85]{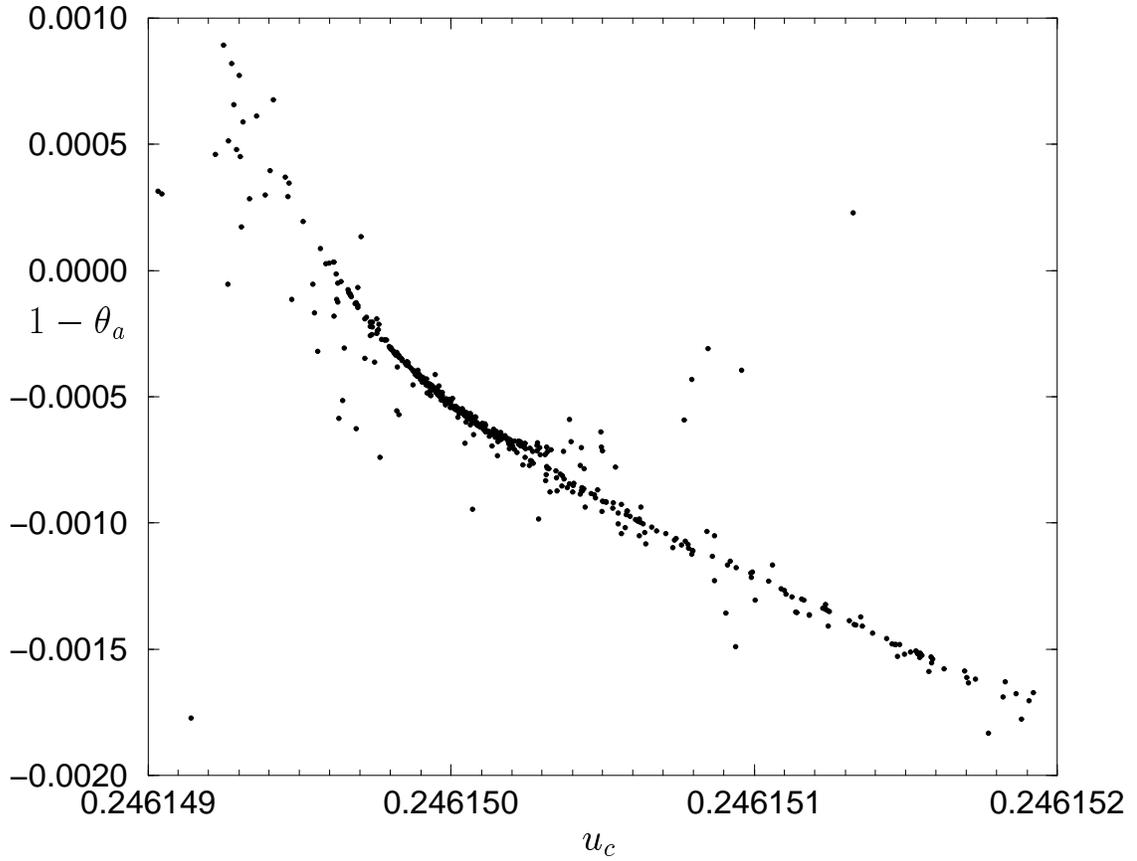}
    
\caption{\label{fig:anicpex}
Estimates for the critical exponent $1-\theta_{\rm a}$ vs. the critical 
point $u_c$ as obtained from second order differential approximants
to the series for the generating function of site animals on the
square lattice.}

\end{center}
\end{figure}

\begin{figure}

\begin{center}

\includegraphics[scale=0.85]{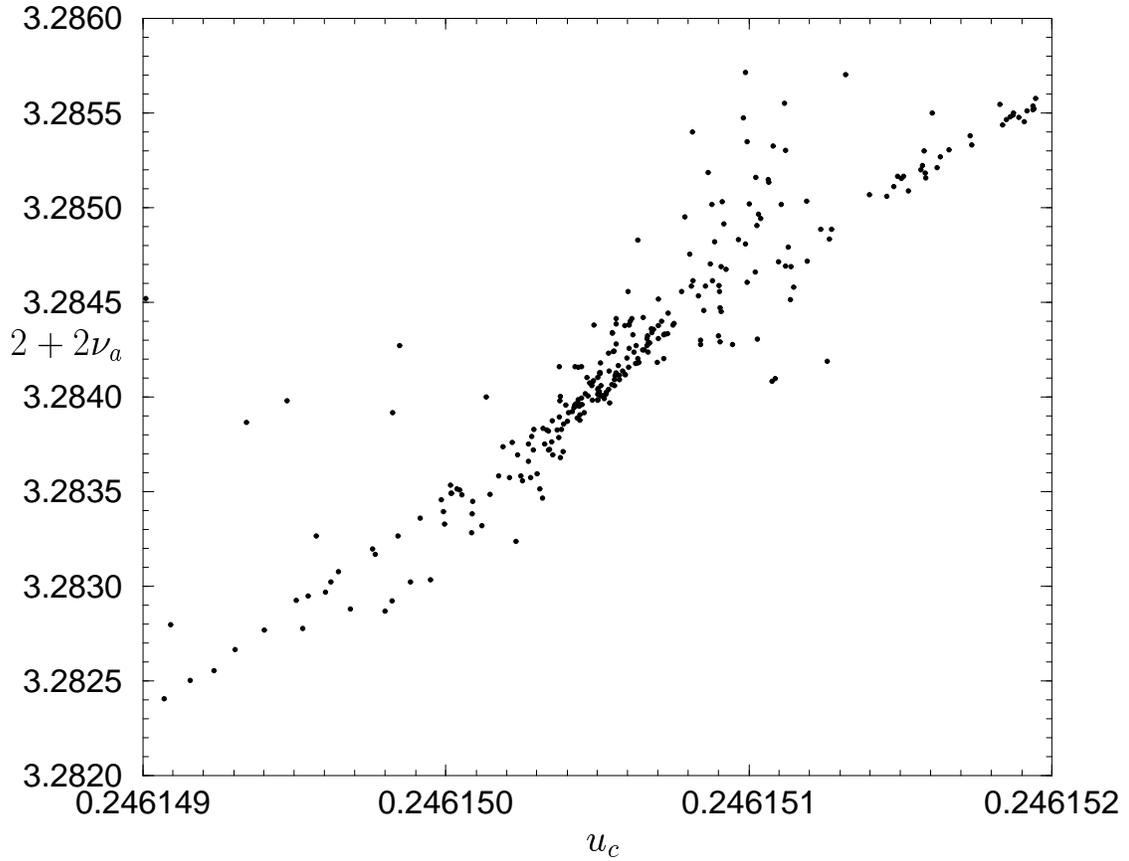}
    
\caption{\label{fig:anirgs}
Estimates for the critical exponent $2+2\nu_{\rm a}$ vs. the critical 
point $u_c$ as obtained from second order differential approximants
to the series for the generating function of the radius of gyration of 
site animals on the square lattice.}

\end{center}
\end{figure}

\end{document}